\documentclass[pdflatex,sn-mathphys-num]{sn-jnl}

\usepackage{graphicx}%
\usepackage{subfigure}
\usepackage{multirow}%
\usepackage{amsmath,amssymb,amsfonts}%
\usepackage{amsthm}%
\usepackage{mathrsfs}%
\usepackage[title]{appendix}%
\usepackage{xcolor}%
\usepackage{textcomp}%
\usepackage{manyfoot}%
\usepackage{booktabs}%
\usepackage{algorithm}%
\usepackage{algorithmicx}%
\usepackage{algpseudocode}%
\usepackage{listings}%

\theoremstyle{thmstyleone}%
\theoremstyle{thmstyletwo}%

\theoremstyle{thmstylethree}%

\begin{document}

\title[Development of the Multichannel Pulsed Ultrasonic Doppler Velocimeter]{Development of the Multichannel Pulsed Ultrasonic Doppler Velocimeter for the measurement of liquid metal flow}


\author[1]{\fnm{Ding-Yi} \sur{Pan}}
\author[1]{\fnm{Yi-Fei} \sur{Huang}}
\author[1]{\fnm{Ze} \sur{Lyu}}
\author*[2]{\fnm{Juan-Cheng} \sur{Yang}}\email{yangjc@xjtu.edu.cn}

\author*[1]{\fnm{Ming-Jiu} \sur{Ni}}\email{mjni@ucas.ac.cn}

\affil[1]{\orgdiv{School of Engineering Science}, \orgname{University of Chinese Academy of Sciences}, 
\city{Beijing}, \postcode{101408}, \country{PR China}}

\affil[2]{\orgdiv{State Key Laboratory for Strength and Vibration of Mechanical Structures, School of Aerospace}, \orgname{Xi'an Jiaotong University}, \city{Xi'an}, \postcode{710049}, \state{Shaanxi}, \country{PR China}}



\abstract{In the present study, by adopting the advantage of ultrasonic techniques, we developed a Multichannel Pulsed Ultrasonic Doppler Velocimetry (MPUDV) to measure the two-dimension-two-component (2D–2C) velocity fields of liquid metal flow. Due to the specially designed Ultrasonic host and post-processing scheme, the MPUDV system can reach a high spatiotemporal resolution of $50$ Hz and $3$ mm in the measurement zone of $192 \times 192$ mm$^{2}$. The flow loop contains a cavity test section to ensure a classical recirculating flow was built to validate the accuracy of MPUDV in velocity field measurement. In the initial phase of the study, water with tracer particles was selected as the working liquid to ensure the velocity field measurements by the well-developed Particle Image Velocimetry (PIV). A comparison of the data obtained from the PIV and MPUDV methods revealed less than 3\% differences in the 2D-2C velocity field between the two techniques during simultaneous measurements of the same flow field. This finding strongly demonstrates the reliability of the MPUDV method developed in this paper. Moreover, the ternary alloy GaInSn, which has a melting point below that of room temperature, was selected as the working liquid in the flow loop to validate the efficacy of the MPUDV in measuring 2D-2C velocity fields. A series of tests were conducted in the cavity test section at varying Reynolds numbers, ranging from $9103$ to $24123$. The measurements demonstrated that the MPUDV could accurately measure the flow structures, which were characterized by a central primary circulation eddy and two secondary eddies in the opaque liquid metal. Furthermore, comprehensive analyses of the velocity data obtained by the MPUDV were conducted. It was found that the vortex center of the primary circulating eddy and the size of the secondary eddies undergo significant alterations with varying Reynolds numbers, indicating the influence of inertial force on the flow characteristics in the recirculating flow. It is therefore demonstrated that the current MPUDV methodology is applicable for the measurement of a 2D-2C velocity field in opaque liquid metal flows.}




\maketitle

\section{Introduction}\label{sec1}
As a good choice for the next generation of power plants, nuclear fusion has received plenty of attention from many countries and international communities. The fusion experimental reactors (e.g. ITER, CFETR) have been established for some fundamental studies to pave the way for commercial fusion reactors\cite{yuanxi2006first}. However, numerous challenges must be overcome before achieving continuous power generation from fusion reactors. The blanket, which is an important component to realizing the tritium breeding, converting energy, effective heat transfer, et al., is crucial for the nuclear fusion system\cite{ihli2008review}. Specifically, the concept of a liquid metal blanket is one of the best selections due to the advantages of flowing liquid metal such as low-pressure operation and high thermal conductivity. Currently, the European Union, the United States and China have carried out designs for liquid metal blankets, such as HCLL (European Union)\cite{aiello2017design}, DCLL (United States)\cite{rapisarda2016conceptual}, and COOL (China)\cite{chen2021conceptual}. However, in the Tokamak system, the strong magnetic field that must be adopted to confine the burning plasma introduces the magnetohydrodynamic (MHD) effects in the liquid metal via the induced Lorentz force when flowing across the magnetic lines. Moreover, the large temperature difference in the liquid metal blanket due to the release of nuclear heat may also influence the liquid metal flow by the introduced buoyancy force. Therefore, the design of a liquid metal blanket requires considering the flow characteristics of liquid metal, particularly under the influence of the MHD effects and large temperature gradients\cite{smolentsev2008characterization,davidson2017introduction}. The corresponding velocity measurement method for the liquid metal flow is necessary. Moreover, measuring the velocity of liquid metal flow is also crucial in various industrial applications, such as in metallurgy and materials processing\cite{davidson1999magnetohydrodynamics}. The high temperature and opacity of liquid metals present unique measurement challenges, making using conventional techniques, such as Laser Doppler Velocimetry (LDV) and Particle Image Velocimetry (PIV), unsuitable. Consequently, measuring velocities within liquid metals has become one of the most formidable challenges in both fundamental studies and industrial applications.   

Several specialized methods have been developed for measuring the liquid metal flow, which can be summarized as invasive and non-invasive methods. As invasive techniques, the force reaction probes and potential difference probes can obtain the local velocity at a measured point while having certain inherent limitations, as outlined by Molokov\cite{molokov2007velocity}. The hot-wire anemometry employs a heated wire to quantify the relationships between the flow velocity and the cooling effect of flowing liquid metal on resistance. However, it is hindered by notable drawbacks, including contamination and corrosion, as well as intricate calibration procedures\cite{comte1976hot}. Among non-invasive techniques, Contactless Inductive Flow Tomography (CIFT) can provide the reconstruction of three-dimensional (3D) flow fields\cite{wondrak2014visualization,wondrak2023three}, but it is not suitable for turbulent flows\cite{stefani2000contactless}. In contrast, dynamic neutron radiography requires manual tracer particles and has yet to be demonstrated in obtaining the three-dimensional velocities of liquid metal flow\cite{vsvcepanskis2017assessment}. In light of the advantages of ultrasonic waves such as the ability to penetrate the opaque liquid metal and the Doppler effects when meeting moving particles, Takeda\cite{takeda1986velocity,takeda1991development,takeda1995velocity} developed the Ultrasonic Doppler velocimetry (UDV) during the 1980s-1990s to measure the velocity distribution along the ultrasonic transmission line in liquid metal flow. Afterward, plenty of developments in the UDV method have been achieved, therefore the UDV method has successfully applied to experimental measurements of various liquid metals, including liquid mercury\cite{takeda1987measurement,mashiko2004instantaneous} and liquid sodium\cite{eckert2002velocity}, liquid gallium\cite{brito2001ultrasonic} and gallium-indium-tin alloy\cite{cramer2004local}. Moreover, the UDV method has been demonstrated to obtain the velocity of magnetic fluids under external magnetic field conditions\cite{kikura1999velocity}. Additionally, the UDV method has also been applied to two-phase flow systems, studying bubble motion behavior and mechanisms\cite{wang2003application,murakawa2005application,murai2006turbulent}, as well as liquid-metal jet driven by bubbles\cite{zhang2007flow}. 

However, the UDV has mostly been used to obtain the 1D velocity in some simple flows such as pipe flow\cite{wiklund2008application,wiklund2012line}. For cases with complex flow, the obtained 1-D velocities are not enough to describe the main characteristics of flow, therefore the two-dimensional velocity distribution is necessary. Franke et al.\cite{franke2010ultrasound} presented a novel pulsed-wave ultrasound Doppler system using an array of 25 transducer elements to investigate liquid metal flow, achieving the two-dimensional flow mapping measurement of liquid metals and further tested the capabilities of this measurement system in the study of unsteady liquid metal flows\cite{franke2013two}. M{\"u}ller et al.\cite{muller2012one} adopted seventeen ultrasonic transducers for the reconstruction of 2D jet flow.  Nauber et al.\cite{nauber2013dual,nauber2013novel,nauber2016ultrasound} introduced a multidimensional Ultrasound Array Doppler Velocimeter (UADV) to obtain 2D-2C velocity fields. Maader et al.\cite{mader2017phased} have combined the pulsed Doppler method with the phased array technique, presented the Phased Array Ultrasound Doppler Velocimeter (PAUDV), confirmed the feasibility of the phased array sensors on measuring the flow field of GaInSn at room temperature. Furthermore, The PAUVP method was applied to measure the two-dimensional velocity field at the outlet of a double-bend pipe where the flow was under turbulent conditions. The accuracy of the PAUVP method has been validated by comparing its measured velocity profiles with those obtained using the well-established Particle Image Velocimetry (PIV) method\cite{shwin2018two}. Recently, Tiwari\cite{tiwari2023ultrasonic} proposed a line selection method for optimizing the position of ultrasonic transducers in 2D flows based on data-driven techniques. However, the currently developed ultrasonic velocity measurement systems suffer from a limitation in the temporal resolution of two-dimensional velocity measurements due to the adoption of a time-division multiplexing (TDM) scheme\cite{thieme2018three}, which is the main purpose of the present study.

A classical flow model is utilized to examine the flow transitions with varying velocities. The selected flow configuration is that of a cavity flow with recirculation. This is employed to demonstrate the effectiveness of the present-developed methodology for the extraction of velocity data. Moreover, cavity flow can reflect changes between various complex flow phenomena at different Reynolds (Re) numbers, such as multi-scale vortices, secondary flow, complex three-dimensional flow, unstable laminar flow, transition flow and turbulent flow.  Accordingly, this configuration represents an optimal validation platform for the precision of measurement systems and a traditional physical model for investigating intricate fluid flows. Experimental studies typically employ lid-driven methods to induce fluid flow within the square cavity, aiming to investigate the variations of internal flow with Re and Spanwise Aspect Ratio (SAR)\cite{pan1967steady,koseff1984visualization,koseff1984lid,koseff1984end,rhee1984flow,prasad1989reynolds}, as well as the transition from laminar steady-state to unsteady flow\cite{aidun1991global,liberzon2011experimental}. Similarly, a tube-driven approach is employed in this study. The experimental measurements of the cavity flow, in conjunction with a comparison between ultrasonic measurements and PIV, serve to validate the precision of the two-dimensional ultrasonic measurement system. Furthermore, experimental observations were conducted to investigate the variations in the GaInSn flow pattern within a square cavity (with an aspect ratio of 1:1 and a spanwise aspect ratio of 0.91:1) at different Re numbers.

The structure of the article is as follows. Sect.\ref{sec2} introduces the measurement system. In Sect.\ref{sec3}, the experimental setup and measurement technique are described. In Sect.\ref{sec4}, the measurement accuracy of the MPUDV system is verified and the velocity measurement results of liquid metal are analyzed. In Sect.\ref{sec5}, the results of the experiment are presented, along with a discussion of their applications.

\section{Measerment system}\label{sec2}
\begin{figure}[h]
	\centering
	\includegraphics[width=1.0\linewidth]{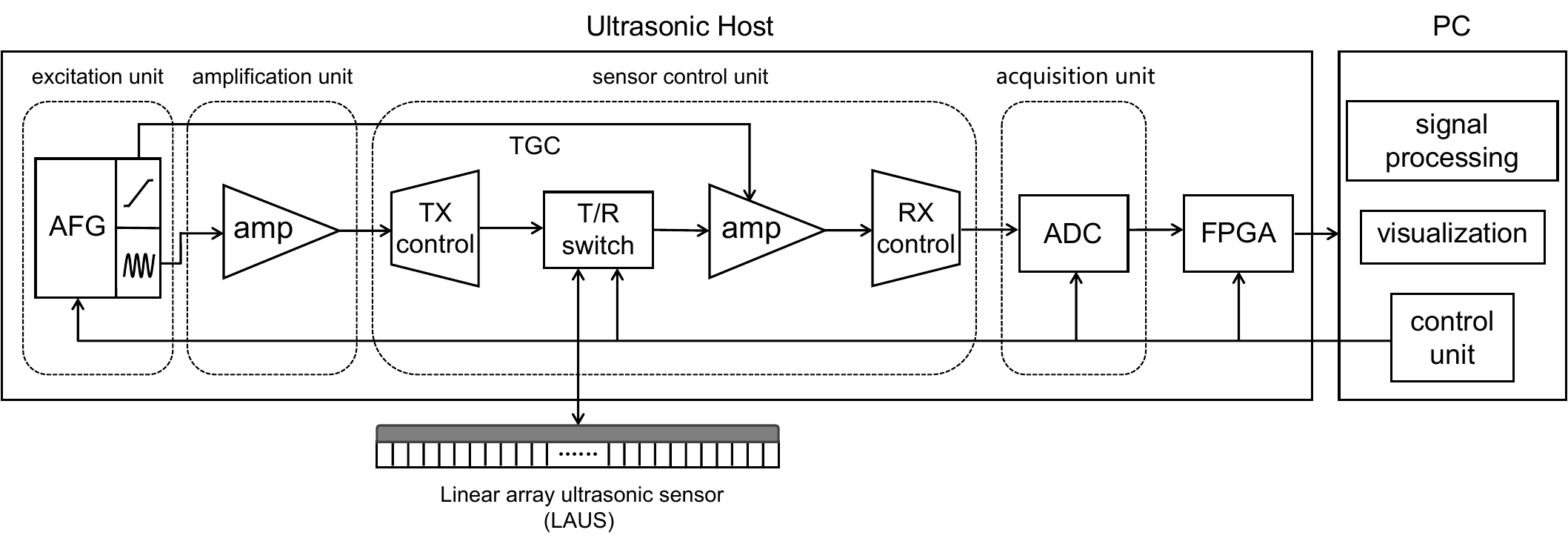}
	\caption{The basic architecture of the MPUDV system.}
	\label{fig:MPUDV}
\end{figure}

The ultrasonic technique is considered an effective method for material tests, including non-destructive testing. It can also be used to measure the flow velocity due to the Doppler effects. In the present study, we develop the Multichannel Pulsed Ultrasonic Doppler Velocimetry (MPUDV) system which consists of ultrasonic hosts, linear array ultrasonic sensors (LAUSs) and the PC. The well-established principles of pulsed ultrasonic Doppler, as outlined by Baker\cite{baker1970pulsed}, were employed to measure fluid flow. Using the MPUDV, the two-dimension-two-component (2D–2C) velocity fields of liquid metal can be realized. The basic architecture of the MPUDV system is shown in figure \ref{fig:MPUDV}, which will be described in more detail in the following sections. 

\subsection{Ultrasonic host}\label{subsec21}
The ultrasound host has 64 discrete operational ultrasound channels with a fast switching time of $5$ ns. As shown in figure \ref{fig:MPUDV}, the ultrasonic host assumes responsibility for the overall control of the system, in addition to the processing of echoes, the computation of velocity fields and their subsequent visualization. The remaining excitation units, amplification units, sensor control units, acquisition units and field-programmable gate array (FPGA) modules collectively constitute the hardware components of the ultrasound host. After configuring channel operation schemes and sending parameter settings to the core control unit of the system, the host PC is responsible for organizing the measurement process. The control unit manages pulse excitation and communication with the PC. Other units simultaneously enter data acquisition and related preparation tasks, thus achieving collaborative coordination among different functional units. The arbitrary function generators (AFG) within the excitation unit generate burst signals for the pulse excitation used to stimulate the ultrasonic sensor. It also controls key ultrasonic parameters, including excitation frequency, pulse repetition frequency (PRF) and other related factors. These signals are subsequently sent to the amplification unit, where the Radio Frequency (RF) amplifier increases the amplitude of the burst signals to a higher voltage level, meeting the requirements for the operation of the sensor's piezoelectric elements. The amplified signals are then transmitted to the sensor control unit which can simultaneously control all ultrasound channels, following the configured channel operation scheme. It controls the amplified pulse excitation transmitted to the respective channels to ensure the flexibility of ultrasound measurements and personalized working schemes. By driving the Transmitting/Receiving (TR) switch, the burst signals received by the corresponding piezoelectric elements of the ultrasonic sensor are separated from the echo signals. It is then amplified using the amplifier with an exponential gain to compensate for the increasing attenuation of the echo signals as they travel through the fluid over time (Time Gain Compensation, TGC). Subsequently, the amplified signal is conveyed through the receive control module to the acquisition unit. The received echo signals by the sensor control unit, after gain and filtering, undergo digitization and recording using the equipped analog-digital converter (ADC) card ($12$-bit, $25$ MSamples/s). Once converted into digital signals, it is then transmitted to the FPGA, where real-time signal processing is conducted using custom algorithms. The processed signal is transmitted to the control unit through optical cables. The control unit then acquires digital IQ-demodulation of Doppler signals\cite{takeda1991development,takeda1995velocity} and conducts offline digital signal processing in MATLAB to obtain the velocity component along a single direction within the flow field. The considerable number of ultrasound channels gives rise to the generation of a substantial quantity of echo data for long-term experimental measurement. This, in turn, places considerable demand on the signal processing module. 

\begin{figure}[h]
	\centering
	\includegraphics[width=0.8\linewidth]{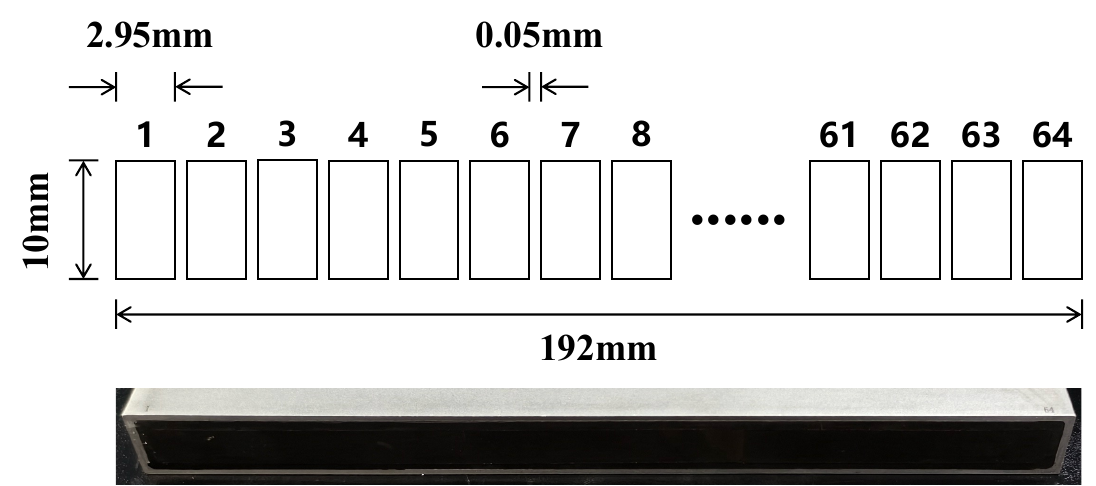}
	\caption{Details of the Linear array ultrasonic sensor (LAUS).}
	\label{fig:LAUS} 
\end{figure}
                                                                               
\subsection{Linear array ultrasonic sensor}\label{subsec22}
The Linear array ultrasonic sensor (LAUS) is an important part of the MPUDV system. Figure \ref{fig:LAUS} illustrates the configuration of LAUS and the corresponding photograph. In contrast to conventional ultrasound Doppler measurement systems, the MPUDV system employs two LAUSs arranged vertically, as shown in Fig.\ref{fig:XYmesh}, wherein each piezoelectric element of a single array sensor functions as both a transmitter for the ultrasound pulses and a receiver for the echo signals. This kind of arrangement allows the realization of one velocity component in a single direction within the two-dimensional flow (2D-1C) at the same time with the help of an ultrasonic host illustrated in \ref{subsec21}. Moreover, to overcome the disturbances from two directional sensors, the MPUDV system uses two LAUSs with emission frequencies of 8MHz and 6MHz, respectively. In figure \ref{fig:LAUS}, each LAUS comprises 64 piezoelectric elements with a size of $2.95 \times 10$ mm$^{2}$, which can be independently operated. The spacing between adjacent piezoelectric elements is $0.05$ mm. As a result, the center spacing of the $64$ measurement lines is $3.0$ mm. Two LAUSs are orthogonally arranged to span the measurement plane of $192 \times 192 $mm$^{2}$. The flow field can be reconstructed by measuring the velocity components at each intersection of the transducer arrays (2D-2C). For piezoelectric elements with the emission frequency $f_{e}$, the number of cycles in a pulse $N_{b}$, the axial resolution in a fluid with a sound velocity of $c$ is expressed by\cite{hedrick1995ultrasound,jensen1996estimation}
\begin{equation}
    \Delta x = \frac{N_b c}{2 f_e}
\end{equation}

Furthermore, the length $L_{n}$ of the near-field region of the sensor can be estimated by
\begin{equation}
    L_{n} = \frac{kw^{2}}{8\lambda}
\end{equation}
Where $k$ is a factor, $w$ is piezoelectric element width ($w = 2.95$ mm), and $\lambda$ is the wavelength of the ultrasonic pulse.

\subsection{Measurement method}\label{subsec23}
The ultrasonic host receives echoes from tracer particles in liquid metal. The echoes contain all the necessary information to reconstruct the velocity profile along the ultrasound beam direction. We assume that the tracer particles follow the fluid flow, therefore the velocity of the particles can reflect the local velocity of the fluid elements. The velocity $v$ along the ultrasound beam can be obtained by estimating the Doppler frequency shift $f_{d}$:
\begin{equation}
    v = \frac{cf_{d}}{2f_{e}}
\end{equation}

The measuring depth $d$ is calculated based on the time offset $t$ between the transmission and reception of ultrasound within the fluid:
\begin{equation}
    d = \frac{tc}{2}
\end{equation}

The echo signal is divided equally into several sections, called gates, to obtain the velocity distribution along the ultrasonic line. Each gate corresponds to an axial distance along the ultrasound beam from the sensor, and the spacing between gates is usually matched with the axial spatial resolution. The corresponding velocity at each gate position can be obtained through the phase shift of the subsequent echo signals, which corresponds to the displacement of scattering particles in the fluid between two consecutive pulses relative to the sensor's position\cite{takeda1995velocity,jensen1996estimation,shung2005diagnostic}.

\begin{figure}[h]
\centering
  \includegraphics[width=1.0\linewidth]{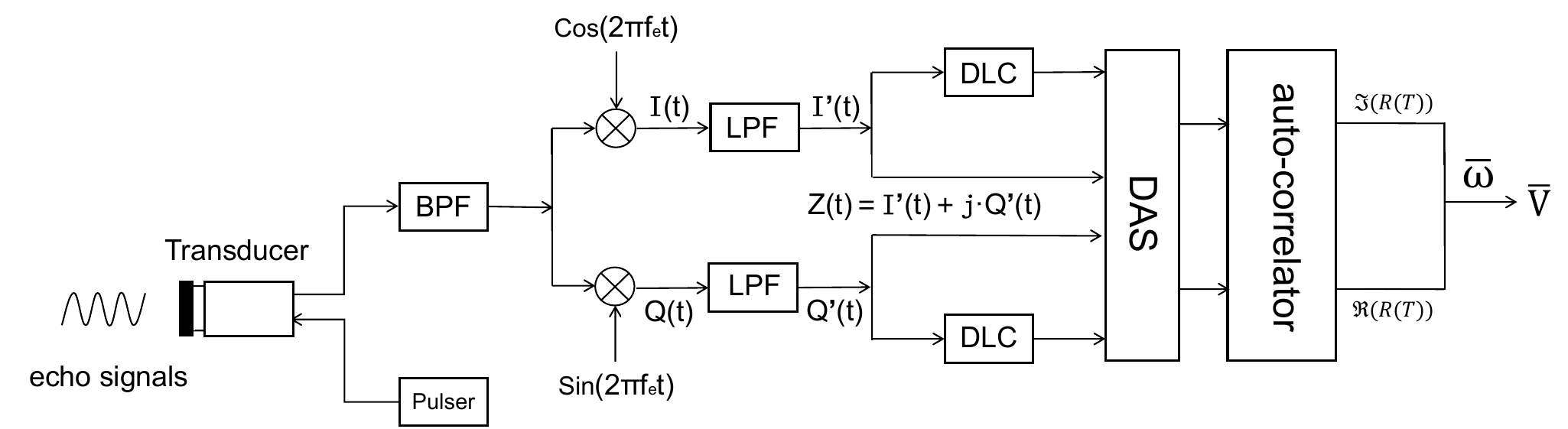}
  \caption{The procedure of the signal process for velocity measurements.}
  \label{fig:Signal_pro}
\end{figure}

The digital signal processing of MPUDV is implemented on both the FPGA module and the PC, thereby enabling the implementation of bespoke pre- and post-processing algorithms, as well as the generation of flow field visualizations following the specific requirements of the measurement task. Figure \ref{fig:Signal_pro} shows the procedure of ultrasonic signal processing. The sampled echo signals are first subjected to bandpass filtering to remove noise, retaining only the signal components near the emission frequency $f_{e}$. 

The filtered echo signals are processed using the IQ-demodulation and then the Delay and Sum (DAS) algorithm\cite{perrot2021so} for beamforming. They are multiplied by $cos(2\pi f_{e} t)$ to obtain the in-phase signal (I-component, $I(t)$). Simultaneously, the bandpass-filtered ultrasound echo signals are multiplied by $sin(2\pi f_{e} t)$ to obtain the quadrature signal (Q-component, $Q(t)$). Afterward, the in-phase signal $I(t)$ and quadrature signal $Q(t)$ are subjected to low-pass filtering separately to eliminate high-frequency components resulting from IQ demodulation. Finally, the velocity estimation is performed using autocorrelation algorithms\cite{jensen1996estimation,kasai1985real}, providing the velocity and direction information along the ultrasound beam. 
According to the theory of UDV, the allowable maximum measured velocity can be estimated by system parameters using the equation below:
\begin{equation}
    v_{max} = \frac{cf_{prf}}{4f_{e}}
\end{equation}

\subsection{Sensor scheme and post-processing}\label{subsec24}
Traditional two-dimensional ultrasonic velocity measurement typically employs a sequential time-division multiplexing (TDM) scheme\cite{nauber2013dual,nauber2013novel,nauber2016ultrasound}. By controlling the minimum spacing and time difference between the piezoelectric elements, and having each piezoelectric element alternately transmit and receive according to the measurement scheme, sound field overlap is avoided to achieve higher spatial resolution. TDM scheme helps reduce ultrasonic crosstalk between piezoelectric elements to a certain extent. However, for the complete two-dimensional flow field, the measurement lines corresponding to velocity components in the same direction are generated in an alternating manner, lacking simultaneity. Furthermore, increasing the number of piezoelectric elements extends the measurement time, causing a significant delay between the first and last measurement lines. This considerable decrease in temporal resolution makes the technique unsuitable for turbulent flows requiring high temporal resolution. 

\begin{figure}[h]
	\centering
	\includegraphics[width=0.8\linewidth]{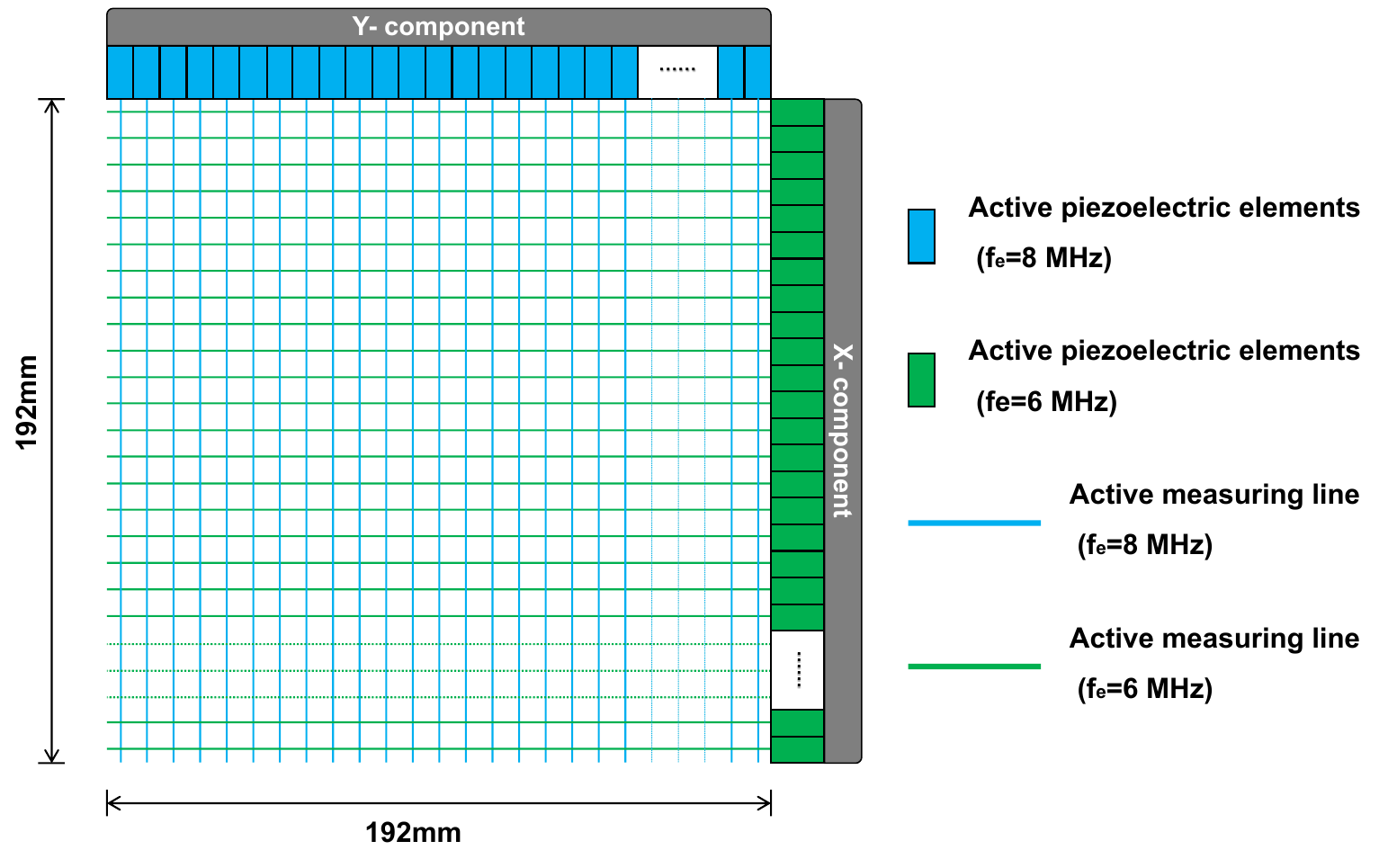}
	\caption{Different-frequency sensor scheme for velocity vector measurements.}
	\label{fig:XYmesh}
\end{figure}

In the present study, as shown in figure \ref{fig:XYmesh}, we can utilize the measurement data from all piezoelectric elements on a single ultrasonic sensor and integrate and process this data to obtain the velocity components along all measurement lines in a single direction by optimizing the algorithm. Therefore, the adjacent piezoelectric elements in the same direction can work simultaneously without influencing one another. On the other hand, by arranging ultrasonic array sensors with different emission frequencies in mutually orthogonally directions, completely distinct ultrasonic signals are produced. This guarantees that upon reception of the echo signal by the sensor, a matching of the ultrasonic frequency is made by the characteristics of the sensor itself. In this way, the alternating operation observed in perpendicular array sensors, as in the sequential time-division multiplexing scheme, is replaced. As a result, all piezoelectric elements in mutually perpendicular ultrasonic arrays work simultaneously. This significantly reduces the time wasted on alternate measurements and greatly improves the temporal resolution, while maintaining the spatial resolution of two-dimensional velocity measurements.

Subsequently, the collected echo data is processed offline using digital signal processing techniques, resulting in the acquisition of the velocity component along a single direction within the flow field. In figure \ref{fig:XYmesh}, it can be observed that two LAUSs are orthogonally arranged to obtain the two-dimensional flow field in the plane. However, ultrasound parameters impact the axial resolution, leading to discrepancies between measurement points along each measurement line and the grid point positions. To address this, a post-processing algorithm was developed to make the necessary adjustments. This algorithm allows for the recombination of a $64 \times 64$ two-dimensional velocity vector field (2D-2C). The resulting vector field combines all the measurement data and is used for flow field visualization. For a specific experimental setup, the positions of all array sensors are fixed and aligned. Subsequently, the measurement volume is overlapped with the sensor positions to form an equidistant grid ($3.0 \times 3.0$ mm$^{2}$). All measurement lines are combined based on their respective geometric positions and orientations. Data interpolation is performed along the axial direction and algorithm parameters are adjusted for accurate velocity synthesis.

\section{Experimental system}\label{sec3}
\begin{figure}[h]
\centering
  \includegraphics[width=1.0\linewidth]{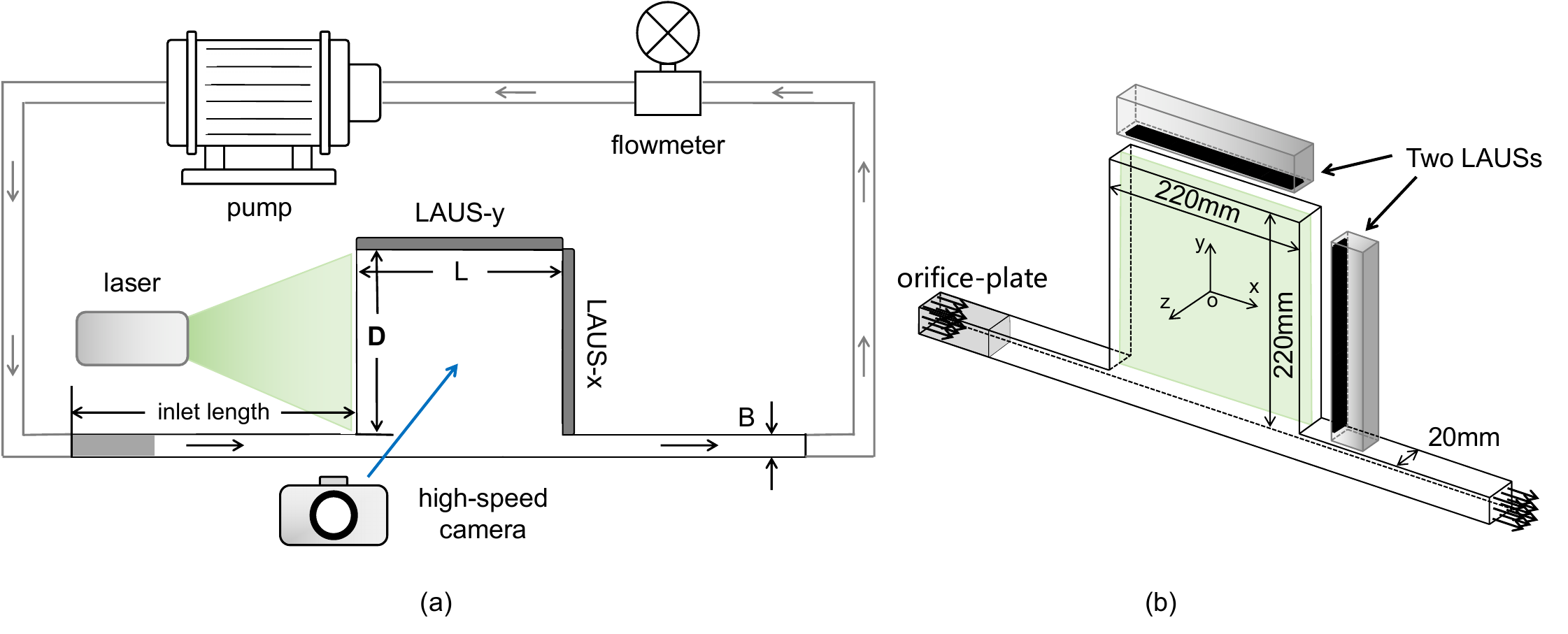}
  \caption{Experimental flow loop system (a) and the test section (b).}
  \label{fig:expsys} 
\end{figure}

To validate the applicability of the MPUDV system, we built an experimental system to measure the velocity distribution in the cavity flow. As shown in figure \ref{fig:expsys}, the experimental system consists of a mechanical pump, flowmeter and a rectangular cavity made of $8$ mm thick acrylic glass as a test cross-section. A circulating flow is kept in the experimental system by the mechanical pump. Regarding the rectangular cavity which is adopted as the test section, it has a lateral width $L = 220$ mm along the bottom tube direction, a vertical height $D = 220$ mm, and a spanwise width $B = 20$ mm, resulting in an aspect ratio of $1:1$ for the rectangular cavity and a spanwise aspect ratio of $0.91:1$. The cross-section of the bottom tube is a $20 \times 20$ mm$^{2}$ rectangle. The test section is connected to two square tubes with a size of $20 \times 20$ mm$^{2}$. In the inlet part, the square tube with a length of $300$ mm is used to generate a uniform jet flow driving the flow inside the test section. Here a honeycomb-like orifice-plate structure of approximately $60$ mm in length is inserted to ensure the full development of the flow inside the square tube. It should be noted that the outlet square tube is only compatible with the flow loop. The mechanical pump is responsible for driving the flow circulation of the entire experimental system, with a flow rate ranging from $0.035$ to $2.700$ L/min. In the flow system, water and GaInSn are used separately as working fluids in this system for different purposes.

\begin{table}[h]
    \centering
    \begin{tabular}{ccc}
        \toprule[1pt]
        \textbf{Parameters} & \textbf{Value (Water)} & \textbf{Value (GaInSn)} \\
        \midrule
        Emission frequency $f_{e}$ [\textit{MHz}] & \multicolumn{2}{c}{6, 8}\\
        Number of cycle in a pulse $N_{b}$ [ - ] & \multicolumn{2}{c}{12, 16}\\
        Pulse repetition frequency $F_{prf}$ [\textit{Hz}] & \multicolumn{2}{c}{1000}\\
        Sound velocity $c$ [\textit{m/s}] & 1480 & 2720 \\
        Number of velocity profiles $N_{vp}$ [ - ] & 2000 & 200 \\
        Temporal resolution $\bigtriangleup t$ [\textit{ms}] & 20 & 200 \\
        Axial resolution $\bigtriangleup x$ [\textit{mm}] & 1.48 & 2.72 \\
        \toprule[1pt]
    \end{tabular}
    \caption{Parameters of the MPUDV system.}
    \label{tab:Tab_1}
\end{table}

The LAUSs arranged in the $X$ and $Y$ directions are mechanically fixed in close contact with the top and sides of the square cavity to measure the internal flow field, see Fig.\ref{fig:expsys}(b). Table \ref{tab:Tab_1} shows the ultrasonic measurement parameters. 

As an initial step, the measurement accuracy of the MPUDV system is validated through the utilization of Particle Image Velocimetry (PIV), which enables the simultaneous acquisition of flow field data. The PIV system consists of a high-speed camera, continuous laser source, and image acquisition system. The high-speed camera used in this system is the Revealer 5KF10-M series, capable of capturing images up to a maximum resolution of $1280 \times 860$ pixels (with a pixel size of $13.7$ $\mu$m) and achieving a maximum frame rate of $3600$ fps. It was positioned directly in front of the measurement area within the cavity. In the experiment, images were captured at a size of $860 \times 860$ pixels, covering an area of $220 \times 220$ mm$^{2}$. The capture rate was set at $50$ fps to match the highest flow rate in the experiment. Additionally, velocity profiles were obtained and processed from 2000 images for each experimental condition, resulting in a total duration of $40$ s. The $532$ nm continuous laser (model SM-SEMI-10M) with a power of $10$ W, was employed. The laser was emitted from the left side of the cavity, directly towards the center, to measure the axial plane of the fluid. The tracer particle model is MV-H0520, primarily composed of glass (a mixture of sodium silicate, sodium carbonate, and silicon dioxide, sintered). Its refractive index is $1.33$, with a particle size of $5-20$ $\mu$m. The density $\rho_{p}=1.05$ g/cm$^{3}$, which is close to that of water, ensures good tracking capability. The image analysis and processing were performed using PIV Lab in Matlab software. By employing autocorrelation analysis on successive frames of PIV images, the mean particle velocity within the interrogation region can be calculated, as can be regarded as the velocity field.

\section{Results and Discussion}\label{sec4}
\subsection{Validation of MPUDV system}\label{subsec41}
We carried out the validation experiments by filling the experimental system shown in figure \ref{fig:expsys} with water which contains tracer particles. According to the previous publication, the flow inside the cavity test section should be in a quasi-two-dimensional (Q2D) state with approximately zero velocity component along the $z$ direction. A change in the flow rate results in a corresponding variation in the Re numbers of the experiments, which range from $9103$ to $24123$. The definition of Re number in the present square cavity is $ (\rho U_{b}L)/\mu$, where $U_{b}$ is the average fluid velocity within the square tube, $L$ is the lateral width of the cavity, $\rho$ is the fluid density, and $\mu$ is dynamic viscosity coefficient of the fluid.

\subsubsection{Visulization of the water flow}\label{subsec411}
\begin{figure}[h]
\centering
  \includegraphics[width=1.0\linewidth]{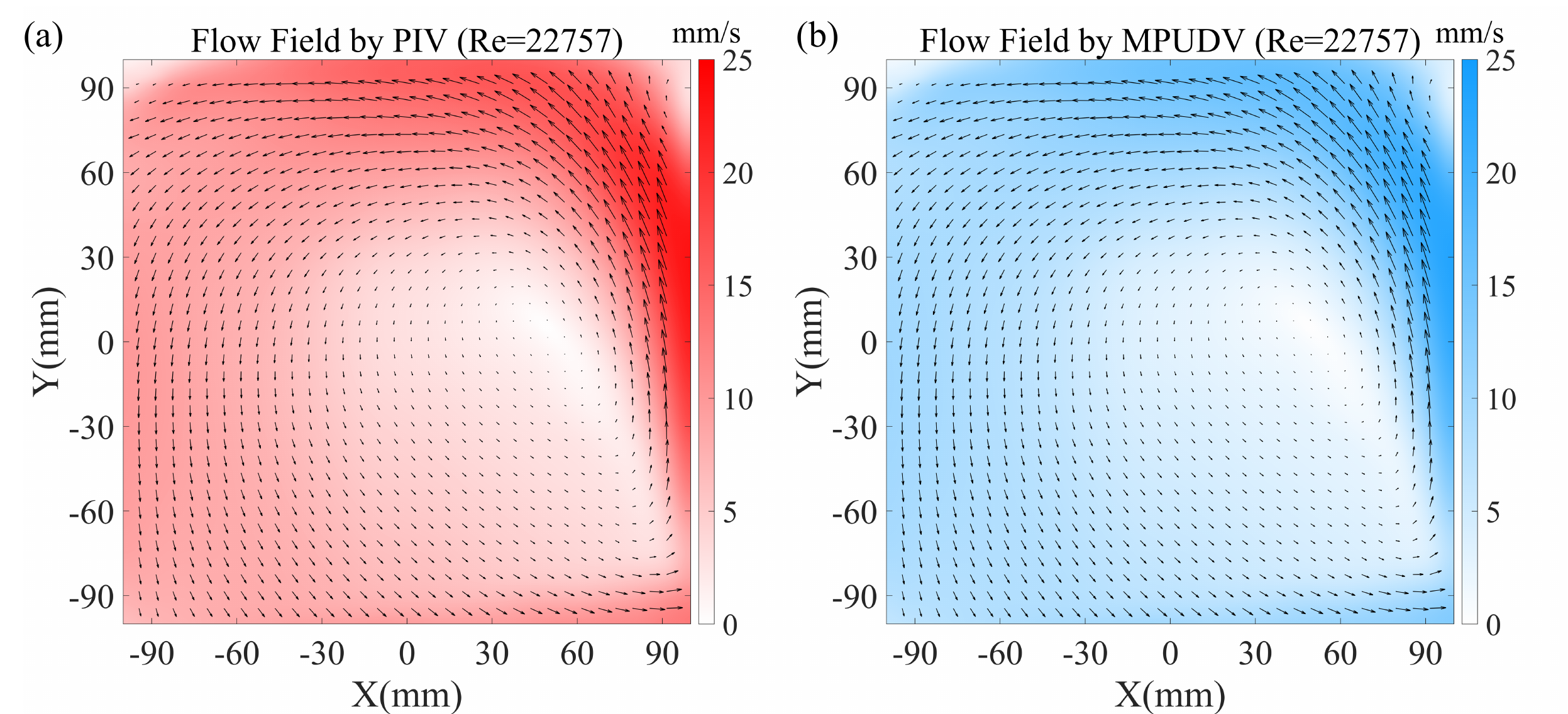}
  \caption{Comparisons of the time-averaged velocity field in the cavity (water) at $Re = 22757$ measured by PIV (a) and MPUDV (b).}
  \label{fig:UDV_PIV} 
\end{figure}

\begin{figure}[h]
	\centering
	\includegraphics[width=0.8\linewidth]{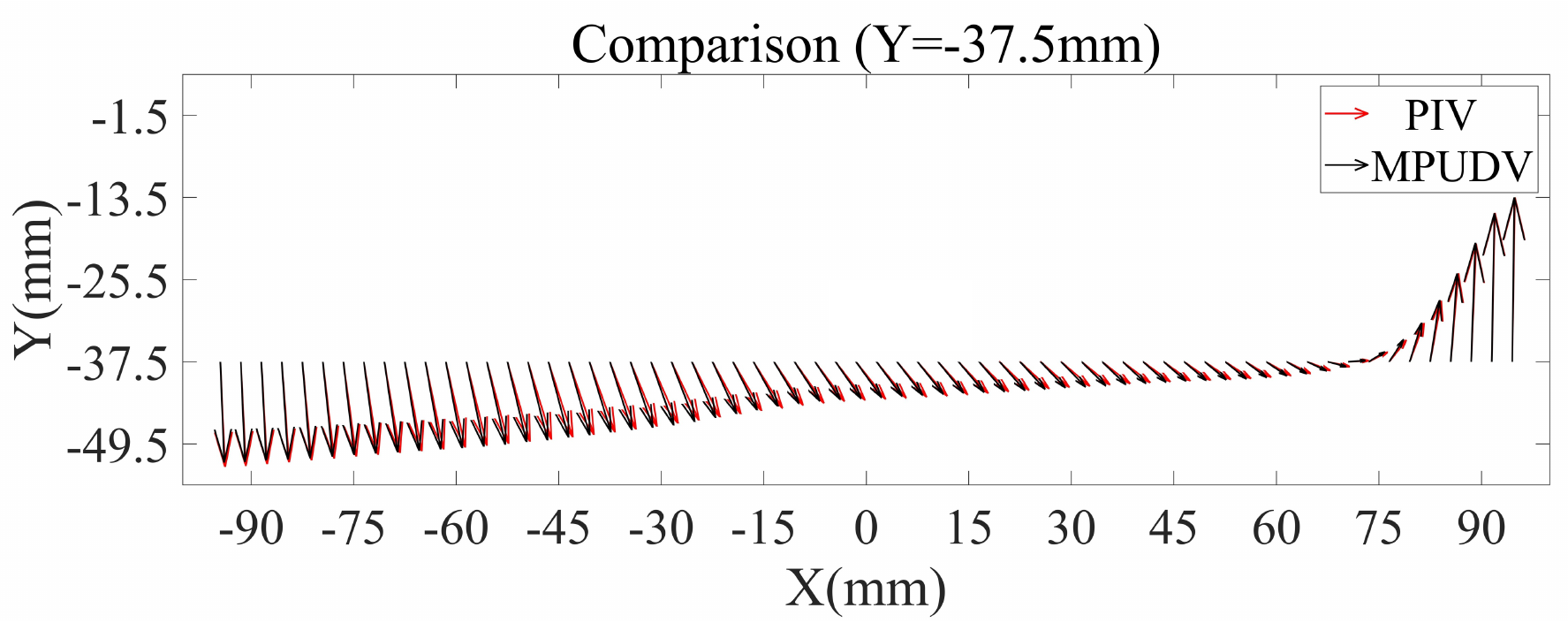}
	\caption{Comparisons of the time-averaged velocity on Line $Y = -37.5$ mm, where the black velocity vector is from MPUDV, the red velocity vector is from PIV.}
	\label{fig:vel_XY} 
\end{figure}

The flow field measurement area is restricted within the range of $-96$ mm $\le X \le$ $96$ mm and $-96$ mm $\le Y \le$ $96$ mm, allowing for a clear observation of the rotating flow within the square cavity. 
Figure \ref{fig:UDV_PIV} illustrates the time-averaged velocity vector field in 2D-2C as measured by two distinct methods: the well-developed PIV method and the MPUDV method developed in the present paper. It can be seen clearly that the velocity vectors obtained by PIV and MPUDV are nearly the same. Furthermore, the velocity vectors distributed along a measured line, $Y = -37.5$ mm, are plotted in figure \ref{fig:vel_XY} to clarify the differences in velocity measurement by PIV and MPUDV. 

Therefore, the experimental results indicate that the two measurement techniques exhibit excellent consistency in velocity measurements of the two-dimensional time-averaged flow field within the square cavity, with good agreement in both magnitude and direction. Furthermore, it can maintain a comparable level of measurement accuracy to PIV. Subsequently, by calculating the relative errors ($\left|V_{MPUDV}-V_{PIV}\right |/V_{MPUDV}$) of velocity measurements at all grid points and then averaging these results, it is demonstrated that the time-averaged velocity measurement error is maintained at less than 3\%.

\subsubsection{Quatitatively comparisons in velocities}\label{subsec412}
\begin{figure}[h]
    \centering
    \includegraphics[width=1.0\linewidth]{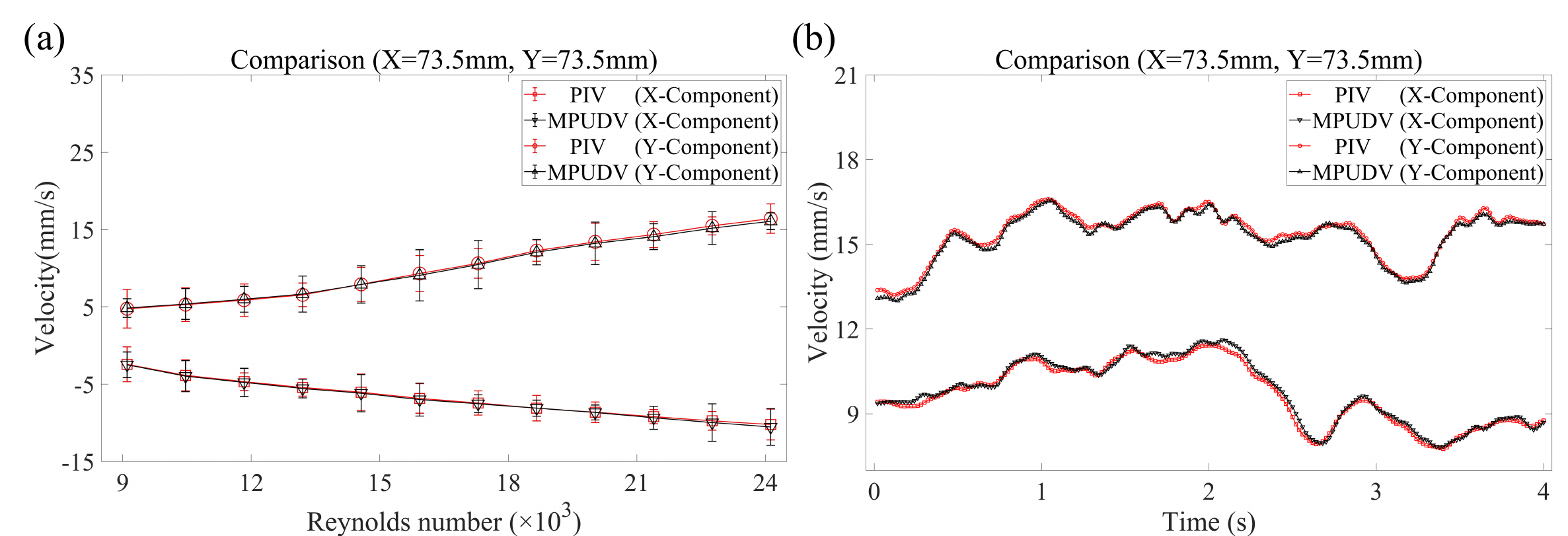}
    \caption{Comparision of velocity data on point $X = 73.5$ mm, $Y = 73.5$ mm, (a) averaged velocity varied with Re number, (b) instantaneous velocity with a time interval of $20$ ms}
    \label{fig:vel_ST_com}
\end{figure}

Since the PIV system and MPUDV system can work simultaneously by the time synchronizer, the comparison of instantaneous velocity is possible in the present study. Thus, to further validate the MPUDV method, we compare the values of velocity varying with Re and time from MPUDV and PIV at a typical measurement point, $X = 73.5$ mm, $Y = 73.5$ mm, displayed in figure \ref{fig:vel_ST_com}. The MPUDV system is capable of accurately capturing the subtle variations in the time-averaged velocity of the square cavity internal flow field at different Re numbers, exhibiting millimeter-level precision. The velocity results obtained from both measurement techniques still exhibit high consistency in their temporal variations, validating the time accuracy of the MPUDV system. Similarly, by calculating and averaging the relative error at each moment ($200$ s in total), compared with the well-established PIV measurement system, the relative measurement error of the MPUDV system is within 3\% at high temporal resolutions, proving for precise measurement of the instantaneous flow field.

\subsection{Liquid Metal Measurement}\label{subsec42}
As illustrated in the previous section, the present MPUDV system can obtain the 2D-2C velocity field of the water flow with high accuracy as compared to the data from the PIV system. Therefore, we have confidence that the MPUVP system can be effectively applied to measuring two-dimensional velocity fields in liquid metals. We then changed the working liquid in the flow system from water to the liquid metal, GaInSn, and systematically measured the liquid metal cavity flow in the same range of $Re = 9103-24123$. The two-dimensional velocity profile inside the square cavity, the size of secondary eddies and the vortex core position with changing Re were obtained and shown in detail in the following subsections. Regarding the tracer particles in liquid metal, the oxide particles are in suspension by nature for the MPUDV measurements.

\subsubsection{Visulization of liquid metal flow}\label{subsec421}
\begin{figure}[h]
\centering
  \includegraphics[width=1.0\linewidth]{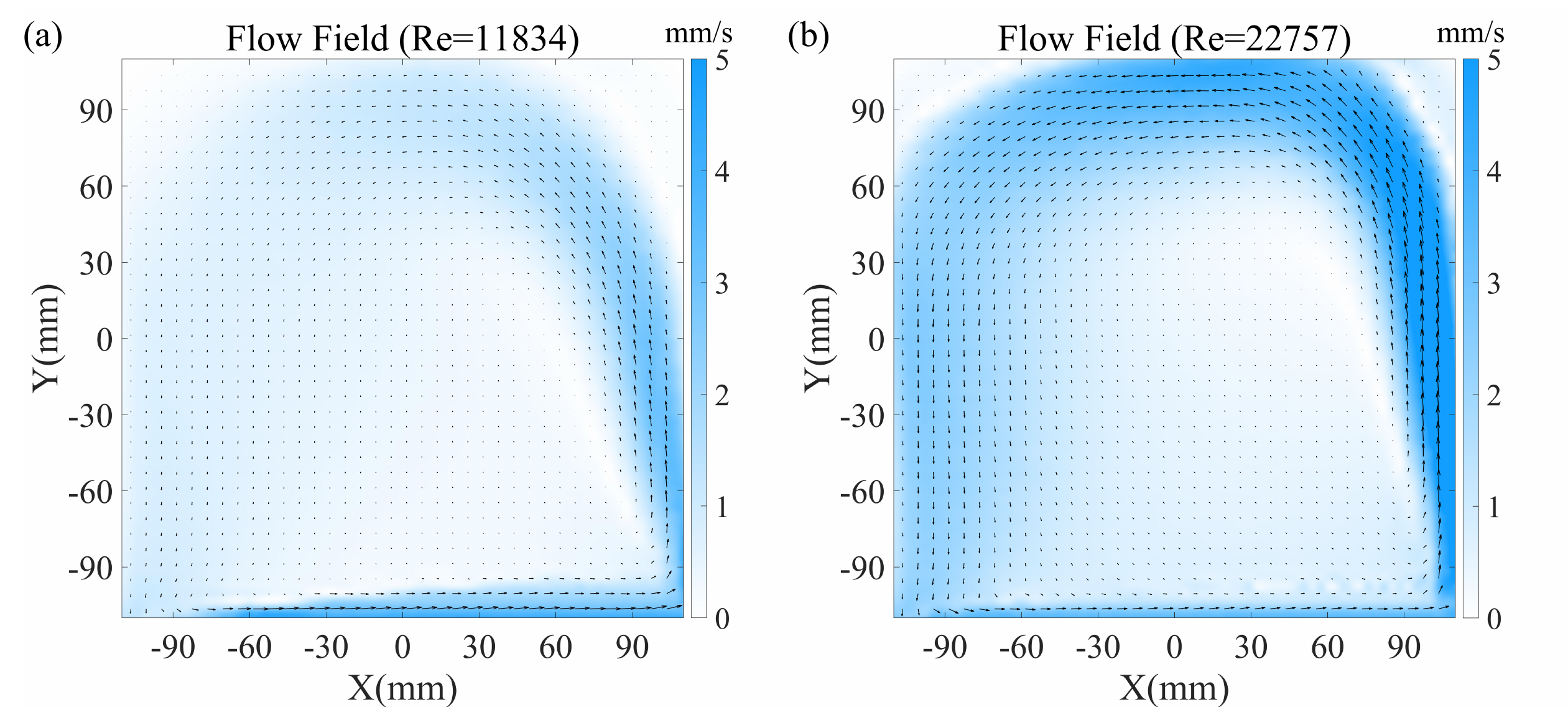}
  \caption{Time-averaged velocity field of the cavity flow (in GaInSn) at $Re=11834$ (a), $Re=22757$ (b).}
  \label{fig:vec_G} 
\end{figure}

The 2D-2C flow field of liquid metal flow measured by MPUDV at two typical Re numbers are plotted in figure \ref{fig:vec_G}. It can be seen that the flow strength is increased with the increase of the Re number, while a large vortex always exists in the cavity. However, it is challenging to discern the secondary eddies in four corners only based on the information presented in figure \ref{fig:vec_G}.       

\begin{figure}[h]
	\centering
	\subfigure{}{\includegraphics[width=0.65\linewidth]{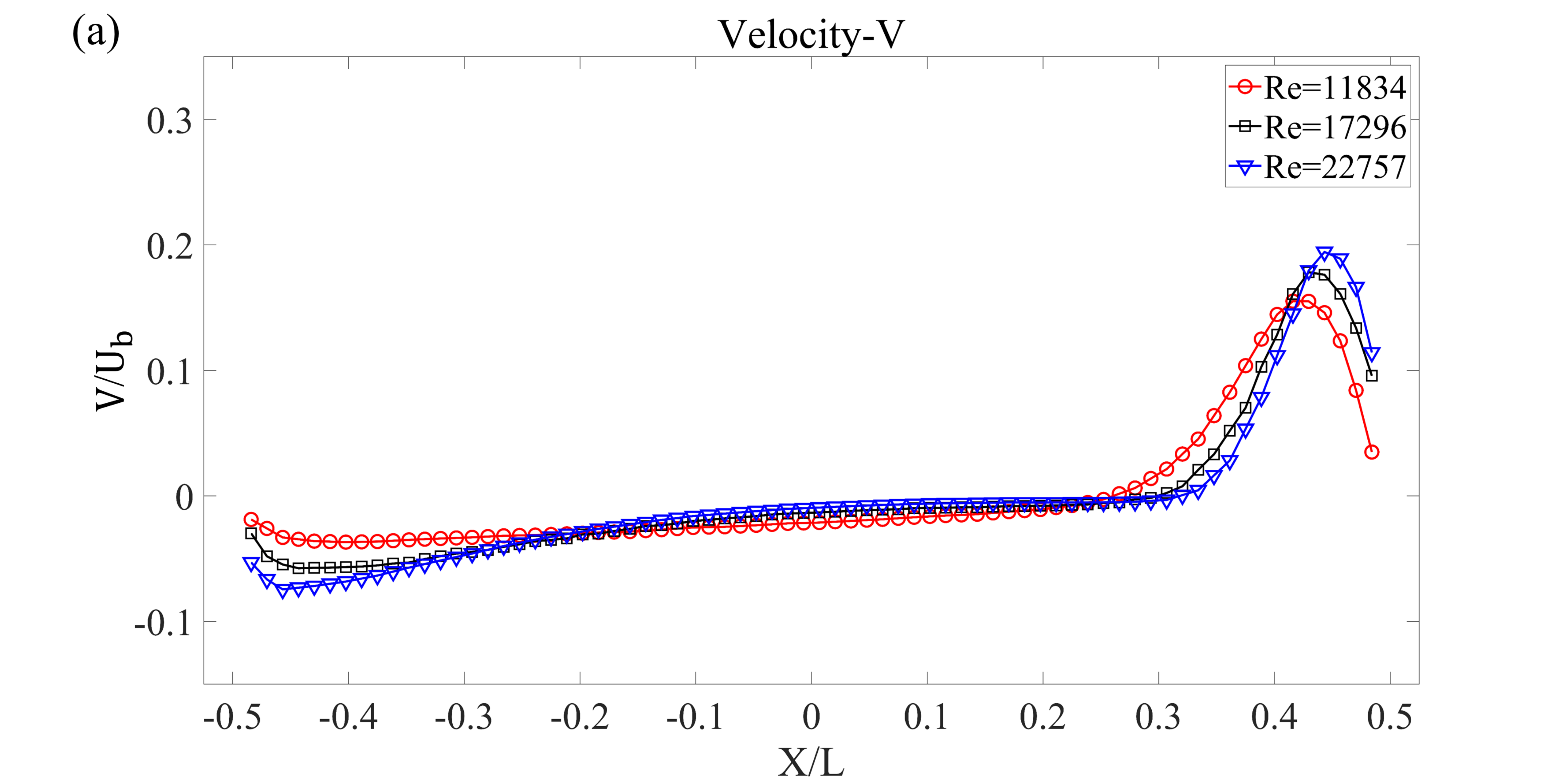}}
	\subfigure{}{\includegraphics[width=0.35\linewidth]{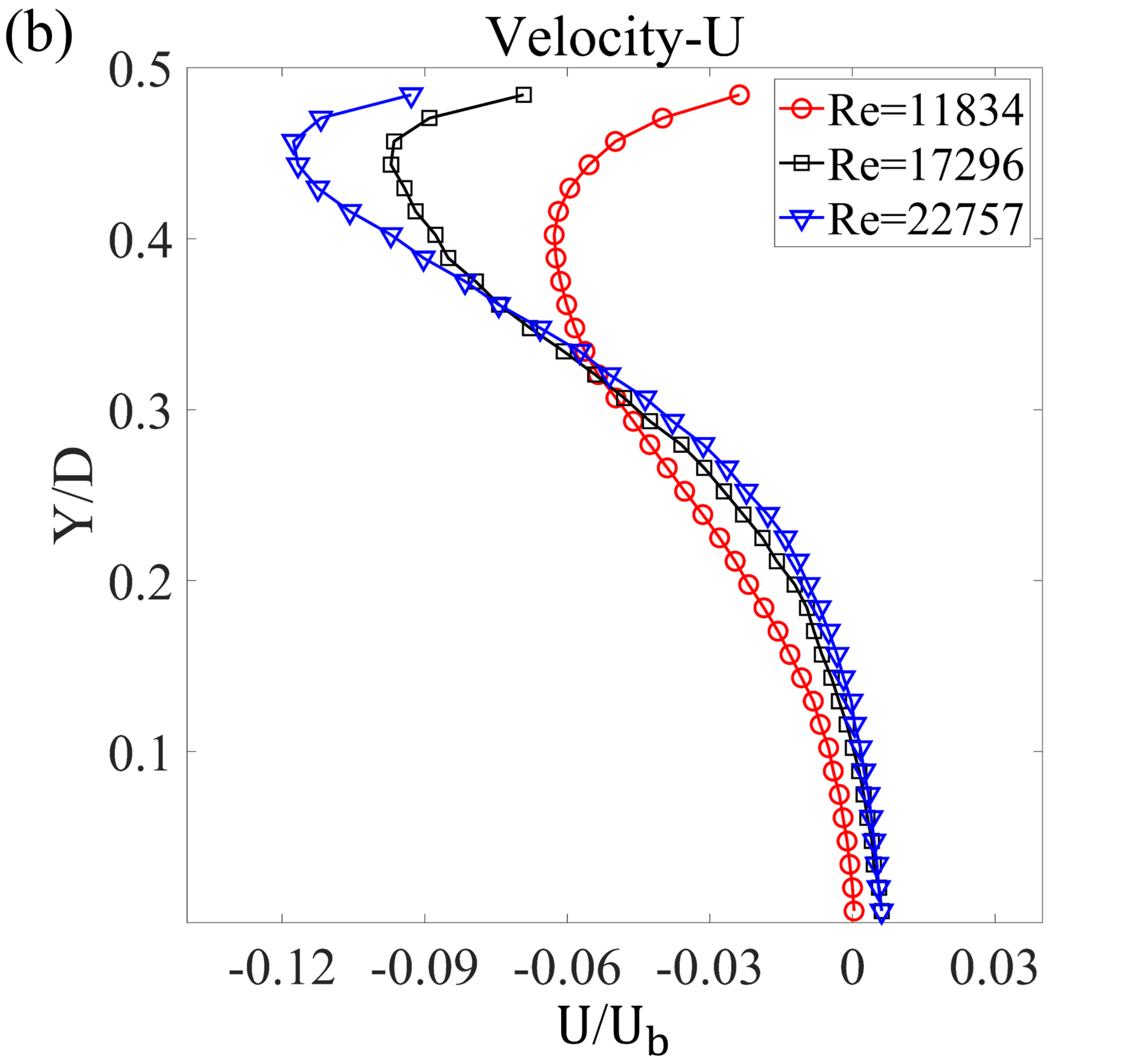}}
	\caption{Mean velocity profile of the horizontal center line (a) and the vertical center line (b) upper portion at $Re = 11834, 17296, 22757$.}
	\label{fig:mean} 
\end{figure}

We then analyze the time-averaged velocity along the horizontal centre line and the upper part of the vertical centre line of the cavity flow field at $Re = 12147, 17296, 22757$, plotted in figure \ref{fig:mean}. The $X$, $Y$ components of velocity were non-dimensionalized using the average flow velocity of the bottom tube $U_{b}$. It can be seen from the velocity distribution that the vortex core remains on the right side of the cavity with higher velocities in the right and top regions of the cavity. With the increase in Re, the formation of secondary eddies leads to noticeable changes in the velocity distribution within the cavity. This is reflected in a lower center velocity along the horizontal center line and higher velocities near the cavity boundaries. The location of the peak velocity on the superior aspect of the vertical centre line undergoes a notable displacement towards the upper boundary of the cavity. Moreover, the overall velocity is susceptible to alteration as a consequence of the displacement of the vortex core. The details of vortex positions are illustrated in the following sections.

\subsubsection{Characteristics of secondary eddy}\label{subsec422}
Regarding the cavity flow, the most striking feature is the variety of secondary eddies when the Re number increases to a certain value. Therefore, we utilized the MPUDV system for corresponding studies. Experimental observations were conducted to study the variations in the size of the Downstream Secondary Eddy (DSE) situated in the upper right region of the cavity and the Upstream Secondary Eddy (USE) in the upper left region. The lateral and longitudinal sizes of secondary eddies were nondimensionalized using the cavity lateral width $L$ and vertical height $D$. The objective of this study was to demonstrate the full range of high-resolution measurement capabilities of the MPUDV system. 

\begin{figure}[H]
	\centering
	\subfigure{}{\includegraphics[width=0.9\linewidth]{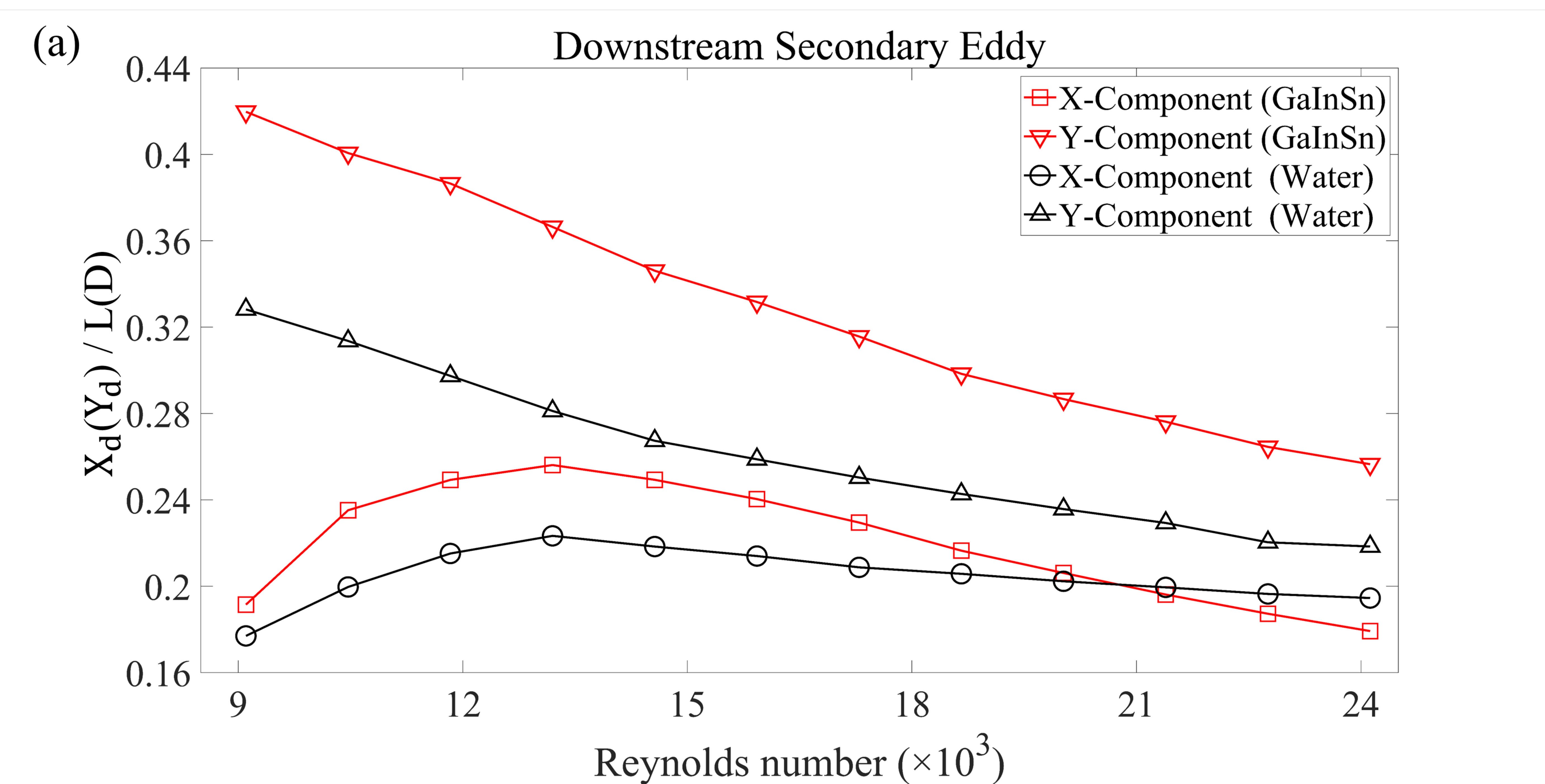}}
	\\
	\subfigure{}{\includegraphics[width=0.9\linewidth]{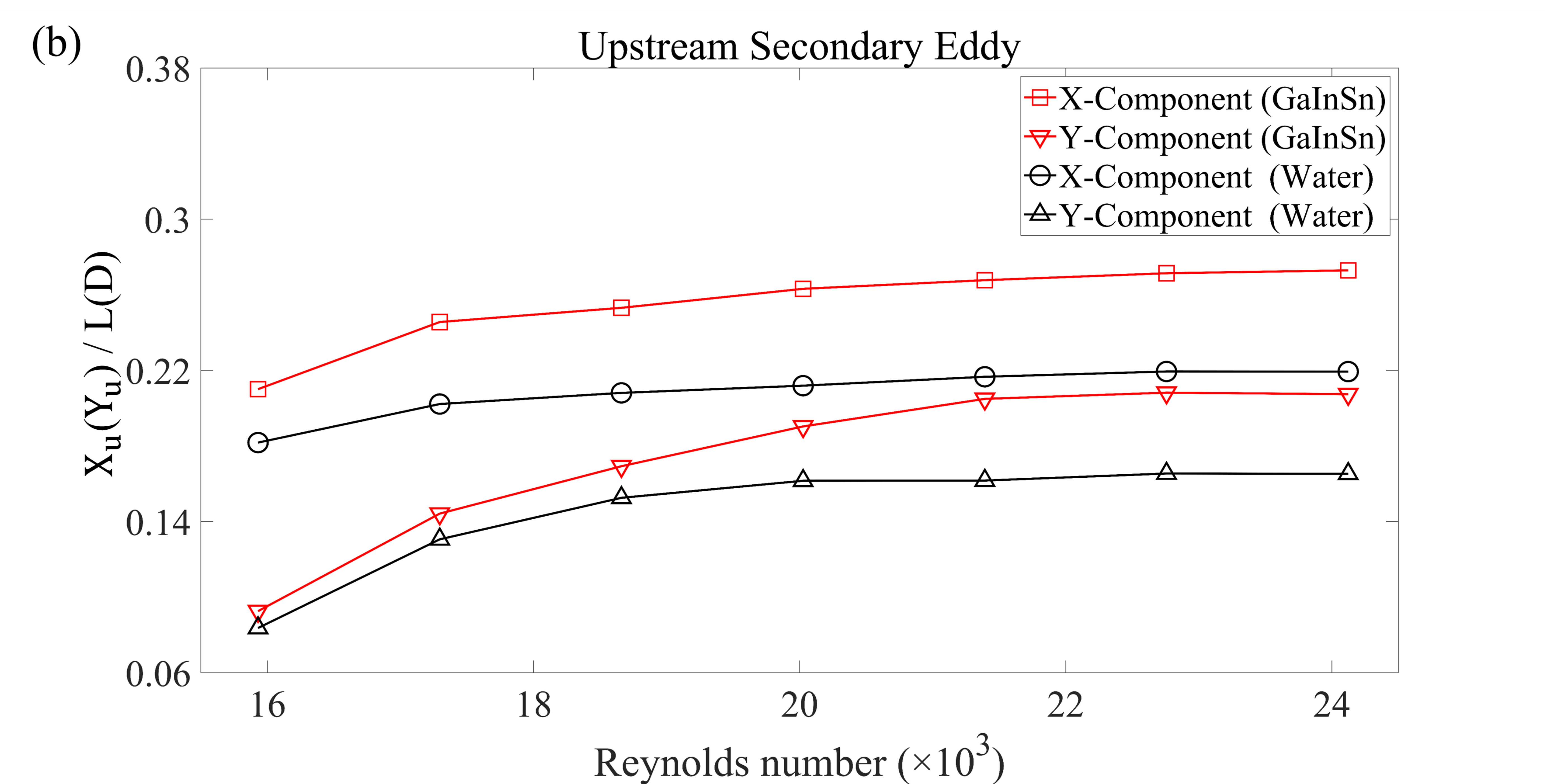}}
	\\
	\subfigure{}{\includegraphics[width=0.9\linewidth]{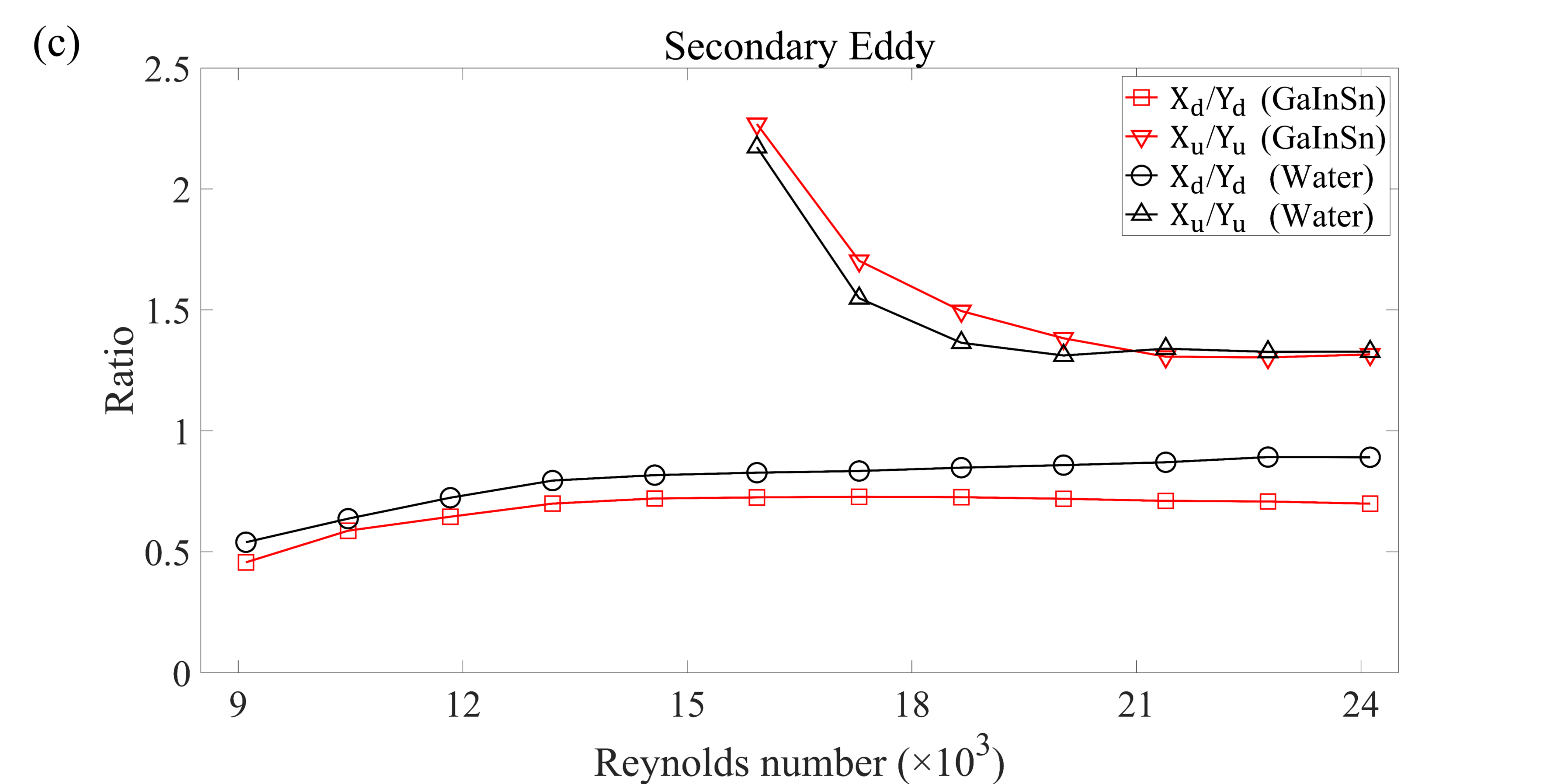}}
	\caption{The size of DSE (a) and USE (b) was measured, and the ratio of the lateral and longitudinal size (c).}
   \label{fig:eddy}
\end{figure}

In figure \ref{fig:eddy}, with the increase of Re, the longitudinal size $Y_{d}$ of the DSE gradually decreases, while the lateral size $Y_{d}$ shows a trend of first increasing and then decreasing. The evolution of both the lateral size $X_{u}$ and the longitudinal size $Y_{u}$ of the USE remains generally consistent. The difference is that the $X_{d}$ of the DSE is smaller than the $Y_{d}$, while it's the opposite for the USE.  Additionally, USE tends to form at higher Re compared to DSE, and the overall variation trends are consistent with the PIV measurement results in water, but due to differences in physical properties, the vortex size in GaInSn is larger, see Fig \ref{fig:eddy}(a), (b). Furthermore, by analyzing the ratio of the lateral and longitudinal sizes of DSE, USE, see Fig. \ref{fig:eddy}(c). In GaInSn and water, the evolution of secondary eddies can be roughly divided into two stages. During the initial stage of gradual formation of the secondary eddies, $X_{d}$ variation of DSE tends to be faster, while USE tends to undergo longitudinal size $Y_{u}$ changes. Subsequently, $\frac{X_{d}}{Y_{d}}$ and $\frac{X_{u}}{Y_{u}}$ remain almost stable, reflecting synchronized changes in the lateral and longitudinal size.

\subsubsection{Displacement of vortex core}\label{subsec423}
\begin{figure}[h]
\centering
  \includegraphics[width=1.0\linewidth]{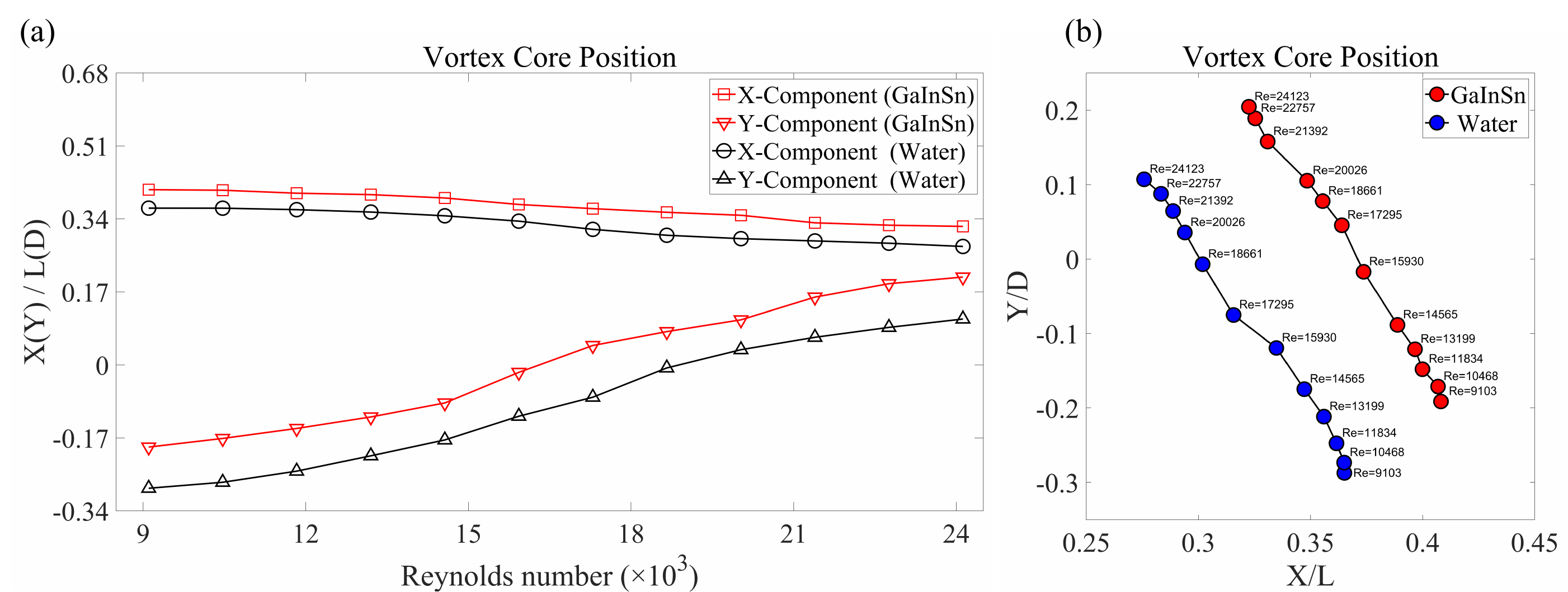}
  \caption{Variation (a) and motion trail (b) of vortex core position (VCP) with Re.}
  \label{fig:trail}
\end{figure}

The evolution of secondary eddies inside the square cavity is closely related to the continuous displacement of the primary circulation eddy. After being non-dimensionalized, the trend of the vortex core position (VCP) of the primary circulation eddy changing with Re was repeatedly measured, see Fig.\ref{fig:trail}(a). An increase in Re leads to a deviation of the vortex center position from the lower-right to the upper-right of the cavity, see Fig.\ref{fig:trail}(b). Starting from Re=14565, due to the gradual formation of USE, the movement speed of the vortex core significantly accelerates for a certain period, resulting in changes in the velocity distribution within the square cavity. While the specific vortex core position in GaInSn and water do not completely coincide, the overall movement patterns remain generally consistent.

\section{Conclusion and outlook}\label{sec5}
In the present study, we developed the Multichannel Pulsed Ultrasonic Doppler Velocimetry (MPUDV) system to obtain the 2D-2C velocity field for opaque liquid metal with high temporal and spatial resolution. The main conclusion can be summarized as follows:

By combining the principles of pulsed ultrasonic Doppler and the array sensors, utilizing two orthogonally linear arranged ultrasonic arrays, it is possible to drive up to 128 piezoelectric elements. This allows for simultaneous measurement of two-dimensional velocity components at $64 \times 64$ grid points across a measurement plane of $192 \times 192$ mm$^{2}$, enabling the capture of flow phenomena and reconstruction of two-dimensional velocity profiles. In comparison to the existing ultrasonic measurement systems, the algorithms and the methodology of operating the sensors at different frequencies simultaneously can achieve a higher temporal resolution. The present MPUDV was validated by the well-developed Particle Image Velocimetry (PIV) in a flow system filled with water. The results demonstrated that the velocity data obtained from the two measurement systems exhibited high consistency, with the MPUDV system demonstrating accuracy within 3\% for both time-averaged velocities and transient changes.  

Furthermore, using the same experimental system with GaInSn as a substitute for water, we apply the MPUDV to measure the 2D-2C velocity field in opaque liquid metal. The experimental validation of cavity flow using GaInSn as the working fluid demonstrated the excellent applicability of MPUDV for measuring liquid metal flows. The experimental results indicate that the flow within the cavity flow exhibits quasi-two-dimensional behavior. With the increase of Re, two secondary eddies (DSE and USE) gradually form within the cavity. The shifting of the central vortex core position leads to different evolution trends in DSE and USE, causing the velocity distribution to become more concentrated towards the boundaries of the cavity. The consistent variation, similar to the PIV measurement results in water, further emphasizes the reliability of the MPUDV system. 

The MPUDV has considerable flexibility in measurement schemes and experimental setups, allowing easy adaptation to various measurement experiments. In addition, the present measurement system has a high temporal resolution, which makes it possible to visualize the liquid metal flow in turbulent flow conditions. To further develop the MPUDV system, It is necessary to extend the operating temperature of ultrasonic sensors to allow measurements in very high-temperature environments. This will be of significant importance for research related to the liquid metal blanket for nuclear fusion.



\bmhead{Acknowledgements}
The authors gratefully acknowledge support from the National Key Research and Development Program of China (no. 2022YFE03130000), NSFC (nos 51927812, 52176089, 52222607), and the Young Talent Support Plan of Xi’an Jiaotong University.









\bibliography{Pan_ref}

\end{document}